# Low Complexity Receiver for Uplink SCMA System via Expectation Propagation

Xiangming Meng, Yiqun Wu, Yan Chen, and Meng Cheng
Huawei Technologies, Co. Ltd., Shanghai, China
Email: {mengxiangming1, wuyiqun, bigbird.chenyan, simon.chengmeng}@huawei.com

*Abstract*—Sparse code multiple access (SCMA) scheme is considered to be one promising non-orthogonal multiple access technology for the future fifth generation (5G) communications. Due to the sparse nature, message passing algorithm (MPA) has been used at the receiver to achieve close to maximum likelihood (ML) detection performance with much lower complexity. However, the complexity order of MPA is still exponential with the size of codebook and the degree of signal superposition on a given resource element. In this paper, we propose a novel low complexity iterative receiver based on expectation propagation algorithm (EPA), which reduces the complexity order from exponential to linear. Simulation results demonstrate that the proposed EPA receiver achieves nearly the same block error rate (BLER) performance as the conventional message passing algorithm (MPA) receiver with orders less complexity.

## I. INTRODUCTION

The future fifth generation (5G) wireless networks are expected to support massive connectivity, low latency as well as better coverage [1]. As one kind of non-orthogonal multiple access (NoMA) technique, sparse code multiple access (SCMA) [2], [3] has been shown to exhibit superior performance and considered to be a promising candidate for 5G multiple access technique. In general, SCMA can be viewed as a generalization of low density spreading (LDS) [4], a sparsely spread code division multiple access (CDMA). Different from LDS, the QAM mapper and the symbol spreader are combined into a single block of SCMA encoder, which directly maps a group of transmitted bits to multidimensional complex domain codewords. As a result, SCMA benefits from the shaping and coding gains of multi-dimensional constellations [5] and thus performs better than LDS with similar decoding complexity. At the receiver side, by leveraging the sparsity of SCMA/LDS codewords, message passing algorithm (MPA) [6] can be applied to achieve near optimal performance.

Despite its excellent performance, SCMA is challenged for its receiver complexity for multi-user detection even with MPA decoder. In general, the order of the complexity of full MPA is exponential with the size of the codebook ($M$) and the degree of signal superposition ($d_f$) on a given resource element, also known as the degree of the function node in the factor graph representation, i.e., $\mathcal{O}(M^{d_f})$. Many efforts have been deployed to either reduce the value of $M$ or $d_f$. Among the most effective ones, low constellation projection in SCMA codebook design can effectively reduce $M$ to $M_p$ ($M_p < M$) for different constellation size, e.g., 4 to 3, 8 to 4, 16 to 9, etc., while giving similar performance compared with no projection [7]. On the other hand, to further reduce complexity, several extensions of MPA have been considered, e.g., Max-log MPA [8], which changes operations from linear domain to log domain to get rid of the exponential calculations, and SIC-MPA [9], which constraints the maximum $d_f$ to $d_s$ ($1 \leq d_s \leq d_f$) by combing with successive interference cancellation (SIC). Note that for $d_s = d_f$, SIC-MPA becomes full MPA, while $d_s = 1$, SIC-MPA reduces to pure SIC receiver. Moreover, most of the aforementioned methods can be applied jointly, e.g., SIC-MPA and Max-log MPA can be combined and then further used together with low projection codebook design. However, even with such combined methods, the receiver still has exponential complexity with respect to the $M_p$ and $d_s$, which may still become impractical for the implementation of very large codebook size (e.g., $M = 64$, $256$, etc.) and very high overload (e.g., $d_f = 8$, $12$, etc.).

In this paper, we investigate another possibility to reduce the complexity of SCMA decoder. Specifically, we consider an uplink coded SCMA system with multiple receive antennas, and propose a low complexity iterative receiver based on expectation propagation algorithm (EPA) [10], [11]. Using EPA, the complexity order of SCMA decoder is reduced to linear complexity, i.e., it only scales linearly with the codebook size $M$ and the average degree of the factor nodes $d_f$ in the factor graph representation. As a result, the computation burden of the SCMA receiver is significantly alleviated and is no longer a problem for implementation in real systems. We shall also show with link level simulation that the proposed EPA receiver achieves nearly the same performance as the full MPA receiver in terms of block error rate (BLER).

In the rest of the paper, we first elaborate on the system model of SCMA transceiver in section II, and then introduce the EPA receiver with comparative complexity analysis in section III. Performance evaluation in terms of BLER is given in section IV, and section V concludes the whole paper.

## II. SYSTEM MODEL

Consider an uplink SCMA system where there are $K$ independent single-antenna users and the base station employs $N_r$ receive antennas, as shown in Fig. 1. Without loss of generality, we assume each user transmits with one SCMA layer. The cases where each user's data takes more than one SCMA layers can be straightforward extension. For each user $k$, the binary information bits $\mathbf{b}_k$ are first encoded by a channel encoder with coding rate $R_k$. To generate SCMA signals, every $J = \log_2 M$ coded bits $\mathbf{c}_k = [c_{k,1}, c_{k,2}, \ldots, c_{k,J}]^T$ are mapped into an $N$-dimensional complex codebook of size $M$, yielding the complex domain SCMA codewords $\mathbf{x}_k = [x_{k,1}, x_{k,2}, \ldots, x_{k,N}]^T$, where $(\cdot)^T$ denotes matrix or vector transpose, and $N$ can be perceived as a spreading factor. The associated mapping function for the $k$-th user is defined as: $f_k: \mathbb{B}^J \to \mathcal{X}_k$, where $\mathcal{X}_k \in \mathbb{C}^N$ is the codebook of user $k$ whose cardinality is $|\mathcal{X}_k| = M$. Note that with the sparse design principle, there are zero elements in $\mathbf{x}_k$ to reduce the collision on each potential physical resource element, as well as to reduce the detection complexity.

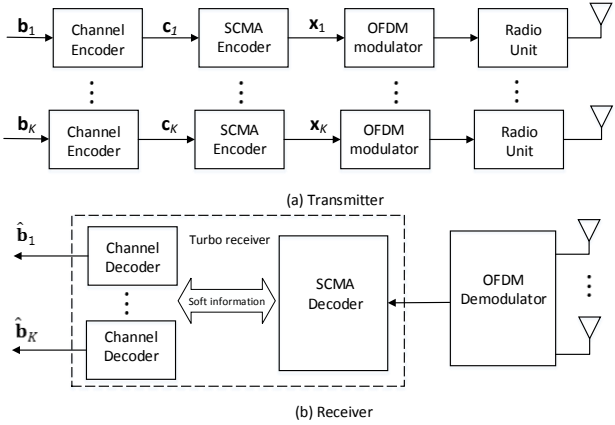

Fig. 1. Block diagram of uplink SCMA system transceiver.

The SCMA codewords $\mathbf{x} = [\mathbf{x}_1^T, \mathbf{x}_2^T, \ldots, \mathbf{x}_K^T]^T$ are then modulated by OFDM modulators. Let $h_{k,n}^{n_r}$ be the channel gain between the user $k$ and the $n_r$-th receive antenna of the base station on the $n$-th resource element, then the corresponding received signal is given by

$$y_n^{n_r} = \sum_{k=1}^K h_{k,n}^{n_r} x_{k,n} + \omega_n^{n_r}, \quad (1)$$

where $\omega_n^{n_r} \sim \mathcal{N}_\mathbb{C}(\omega_n^{n_r}; 0, \sigma^2)$ is the additive noise, and $a \sim \mathcal{N}_\mathbb{C}(a; \mu, \nu)$ indicates that the random variable $a$ follows the complex Gaussian distribution with mean $\mu$ and variance $\nu$. In matrix form, we have

$$\mathbf{y}^{n_r} = \sum_{k=1}^K \mathbf{H}_k^{n_r} \mathbf{x}_k + \boldsymbol{\omega}^{n_r}, \quad (2)$$

where $\mathbf{y}^{n_r} = [y_1^{n_r}, y_2^{n_r}, \ldots, y_N^{n_r}]^T$, $\boldsymbol{\omega}^{n_r} = [\omega_1^{n_r}, \omega_2^{n_r}, \ldots, \omega_N^{n_r}]^T$, $\mathbf{H}_k^{n_r} = \text{diag}(h_{k,1}^{n_r}, h_{k,2}^{n_r}, \ldots, h_{k,N}^{n_r})$, and $\text{diag}(a_1, \ldots, a_L)$ denotes a diagonal matrix with diagonal elements $a_1, \ldots, a_L$.

Consider an iterative multi-user receiver [12] for such SCMA system. Given the received signal $\mathbf{y} = [\mathbf{y}^1; \mathbf{y}^2; \ldots; \mathbf{y}^{N_r}]$ and the channel knowledge $\mathbf{H}_k^{n_r}$, $n_r = 1, \ldots, N_r, k = 1, \ldots, K$, the task of SCMA decoder is to generate extrinsic log likelihood ratio (LLR) for each coded bit of each user, i.e.,

$$\lambda_1(c_{k,j}) = \log \frac{P(c_{k,j}=1|\mathbf{y})}{P(c_{k,j}=0|\mathbf{y})} - \log \frac{P(c_{k,j}=1)}{P(c_{k,j}=0)}$$
$$= \Lambda(c_{k,j}) - \lambda_2^p(c_{k,j}), \quad (3)$$

where $\lambda_2^p(c_{k,j}) = \log \frac{P(c_{k,j}=1)}{P(c_{k,j}=0)}$ denotes the a priori LLR of $c_{k,j}$ given by the channel decoder in the previous iteration. For the first iteration, no prior information is available and thus $\lambda_2^p(c_{k,j}) = 0$. Since the calculation of $\lambda_2^p(c_{k,j})$ has been well studied, the focus of this paper lies in calculating the posterior LLRs, i.e. $\Lambda(c_{k,j}) = \log \frac{P(c_{k,j}=1|\mathbf{y})}{P(c_{k,j}=0|\mathbf{y})}$. Define
$$\mathcal{C}_{k,j}^+ \triangleq \{(c_{k,1}, \ldots, c_{k,j-1}, 1, c_{k,j+1}, \ldots, c_{k,J}): c_{k,i} \in \{0,1\}, i \neq j\}$$
$$\mathcal{C}_{k,j}^- \triangleq \{(c_{k,1}, \ldots, c_{k,j-1}, 0, c_{k,j+1}, \ldots, c_{k,J}): c_{k,i} \in \{0,1\}, i \neq j\}$$

Let $\mathcal{X}_{k,j}^+$ be the set of SCMA codewords mapped from the coded bits in $\mathcal{C}_{k,j}^+$ and similarly for $\mathcal{X}_{k,j}^-$. Then, using the Bayes' rule, we have

$$\Lambda(c_{k,j}) = \log \frac{\sum_{\mathbf{x}_k \in \mathcal{X}_{k,j}^+} P(\mathbf{x}_k|\mathbf{y})}{\sum_{\mathbf{x}_k \in \mathcal{X}_{k,j}^-} P(\mathbf{x}_k|\mathbf{y})} \quad (4)$$

However, direct computation of (4) has a prohibitively high complexity. Leveraging the sparsity of SCMA codewords, message passing algorithm (MPA) and some extensions have been proposed to calculate (4). Nevertheless, MPA still has exponential complexity which is unrealistic in practice, especially for large codebook size and high overload.

III. EPA BASED ITERATIVE RECEIVER

In this section, a low-complexity iterative receiver is proposed based on expectation propagation algorithm (EPA). EPA is an approximate Bayesian inference method in machine learning for estimating sophisticated posterior distributions with simple distributions through distribution projection[10], [11]. Mathematically, the projection of a particular distribution $p$ into some distribution set $\Phi$ is defined as
$$\text{Proj}_\Phi(p) = \arg\min_{q \in \Phi} D(p||q),$$
where $D(p||q)$ denotes the Kullback-Leibler (KL) divergence. If $p \in \Phi$, then the projection reduces to an identity mapping. However, in general, $p \notin \Phi$ and thus the distribution projection is a nonlinear operation.

The joint distribution $P(\mathbf{x}, \mathbf{y})$ can be factorized as

$$P(\mathbf{x}, \mathbf{y}) = P(\mathbf{y}|\mathbf{x}) P(\mathbf{x}) = \prod_{n_r=1}^{N_r} \prod_{n=1}^{N} P(y_n^{n_r}|\mathbf{x}) \prod_{k=1}^{K} P(\mathbf{x}_k) \quad (5)$$

where

$$P(y_n^{n_r}|\mathbf{x}) = \frac{1}{\pi\sigma^2} e^{-\frac{|y_n^{n_r} - \sum_{k=1}^K h_{k,n}^{n_r} x_{k,n}|^2}{\sigma^2}}.$$

A. Factor Graph Representation

The factorization of the joint distribution $P(\mathbf{x}, \mathbf{y})$ in (5) can be represented by a factor graph $\mathcal{G}(\mathcal{V}, \mathcal{N})$, which contains $K$ variable nodes and $N_r N$ likelihood factor nodes (FNs) $f_n^{n_r}$, and $K$ prior factor nodes $\Delta_k$, as shown in Fig. 2.

The variable nodes (VNs) represent data layers and the likelihood FNs $f_n^{n_r}$ represent the likelihood function $P(y_n^{n_r}|\mathbf{x})$, and the prior FN $\Delta_k$ represents the prior probability function $P(\mathbf{x}_k)$, which can be derived from the extrinsic LLRs fed back from the decoders as

$$P(\mathbf{x}_k) = \prod_{j=1}^{J} \frac{e^{c_{k,j} \lambda_2^p(c_{k,j})}}{1 + e^{c_{k,j} \lambda_2^p(c_{k,j})}}, \quad (6)$$

where $c_{k,j}$ is the $j$-th coded bit corresponding to $\mathbf{x}_k$.

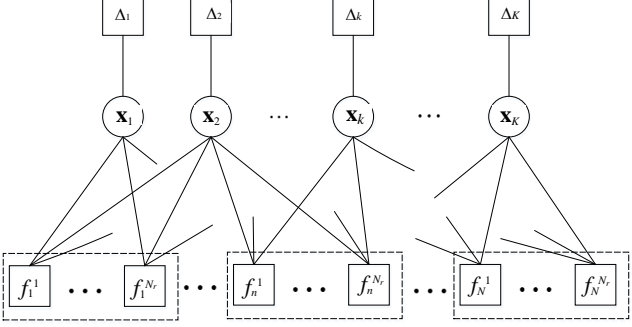

Fig. 2. Factor graph representation of $P(\mathbf{x}, \mathbf{y})$.

The edges between VNs and likelihood FNs in Fig. 2 indicate that the corresponding data layers have nonzero signatures on the associated resources. Denote by $(n, n_r)$ the $n$-th likelihood FN node of receive antenna $n_r$. For ease of notation, let $V(k) = \{n: x_{k,n} \neq 0\}$ denote the neighboring FNs of VN $k$ for each receive antenna, and $F(n) = \{k: x_{k,n} \neq 0\}$ denote the neighboring VNs of the FN $(n, n_r)$ for each receive antenna. The cardinalities of $V(k)$ and $F(n)$ are assumed to be $|V(k)| = d_v, k = 1, \dots, K$, $|F(n)| = d_f, n = 1, \dots, N$, respectively.

### B. EPA SCMA Decoding

Given the factor graph representation, the posterior belief of each symbol can be calculated via EPA. Specifically, let $I_{k \to n, n_r}^t(\mathbf{x}_k)$ be the message from VN node $k$ to FN node $(n, n_r)$ at the $t$-th iteration and $I_{n, n_r \to k}^t(\mathbf{x}_k)$ the message of the opposite direction, respectively. Then, according to the expectation propagation principle [11], the message update rule can be written as

$$I_{k \to n, n_r}^t(\mathbf{x}_k) = \frac{\text{Proj}_\Phi(p^t(\mathbf{x}_k))}{I_{n, n_r \to k}^t(\mathbf{x}_k)}, \quad (7)$$

$$I_{n, n_r \to k}^t(\mathbf{x}_k) = \frac{\text{Proj}_\Phi(q_{n,n_r}^t(\mathbf{x}_k))}{I_{k \to n, n_r}^t(\mathbf{x}_k)}, \quad (8)$$

where

$$p^t(\mathbf{x}_k) = I_{\Delta \to k}(\mathbf{x}_k) \prod_{n_r=1}^{N_r} \prod_{m \in V(k)} I_{m, n_r \to k}^{t-1}(\mathbf{x}_k), \quad (9)$$

$$q_{n,n_r}^t(\mathbf{x}_k) = I_{k \to n, n_r}^t(\mathbf{x}_k) \sum_{\mathbf{x}_l, l \in F(n), l \neq k} f_n^{n_r}(\mathbf{x}) \prod_{l \in F(n), l \neq k} I_{l \to n, n_r}^t(\mathbf{x}_l), \quad (10)$$

and $I_{\Delta \to k}(\mathbf{x}_k) = P(\mathbf{x}_k)$ as defined in (7) and

$$f_n^{n_r}(\mathbf{x}) = \frac{1}{\pi \sigma^2} e^{-\frac{|y_n^{n_r} - \sum_{l \in F(n)} h_{l,n}^{n_r} x_{l,n}|^2}{\sigma^2}}. \quad (11)$$

The main computation burden lies in (10) and is of the order $\mathcal{O}(M^{d_f})$, which is the same as the conventional MPA. To reduce the computational complexity, choosing the projection set $\Phi$ as the complex Gaussian distribution, the message passing then reduces to mean and variance parameter update. Since $\mathbf{x}_k$ is a complex multidimensional vector, it seems that after projection the messages are complex vector Gaussian distributions. However, note from (11) that the $n$-th likelihood factor node only has relationship with $x_{k,n}, k \in F(n)$. As a result, the messages in the update equation (7) - (10) can be simplified to scalar complex Gaussian distributions. Specifically, we have

$$I_{k \to n, n_r}^t(x_{k,n}) = \frac{\text{Proj}_\Phi(p^t(x_{k,n}))}{I_{n, n_r \to k}^t(x_{k,n})}, \quad (12)$$

$$I_{n, n_r \to k}^t(x_{k,n}) = \frac{\text{Proj}_\Phi(q_{n,n_r}^t(x_{k,n}))}{I_{k \to n, n_r}^t(x_{k,n})}. \quad (13)$$

After some algebra, we obtain

$$\text{Proj}_\Phi(p^t(x_{k,n})) \propto \mathcal{N}_{\mathbb{C}}(x_{k,n}; \mu_{k,n}^t, \xi_{k,n}^t) \quad (14)$$
$$I_{k \to n, n_r}^t(x_{k,n}) \propto \mathcal{N}_{\mathbb{C}}(x_{k,n}; \mu_{k \to n, n_r}^t, \xi_{k \to n, n_r}^t) \quad (15)$$
$$I_{n, n_r \to k}^t(x_{k,n}) \propto \mathcal{N}_{\mathbb{C}}(x_{k,n}; \mu_{n, n_r \to k}^t, \xi_{n, n_r \to k}^t) \quad (16)$$

where $\propto$ denotes equality up to scale and

$$\xi_{k \to n, n_r}^t = \left( \frac{1}{\xi_{k,n}^t} - \frac{1}{\xi_{n, n_r \to k}^{t-1}} \right)^{-1} \quad (17)$$

$$\mu_{k \to n, n_r}^t = \xi_{k \to n, n_r}^t \left( \frac{\mu_{k,n}^t}{\xi_{k,n}^t} - \frac{\mu_{n, n_r \to k}^{t-1}}{\xi_{n, n_r \to k}^{t-1}} \right) \quad (18)$$

$$\mu_{n, n_r \to k}^t = \frac{1}{h_{k,n}^{n_r}} \left( y_n^{n_r} - \sum_{l \in F(n), l \neq k} h_{l,n}^{n_r} \mu_{l \to n, n_r}^t \right) \quad (19)$$

$$\xi_{n, n_r \to k}^t = \frac{1}{|h_{k,n}^{n_r}|^2} \left( \sigma^2 + \sum_{l \in F(n), l \neq k} |h_{l,n}^{n_r}|^2 \xi_{l \to n, n_r}^t \right) \quad (20)$$

The posterior mean $\mu_{k,n}^t$ and variance $\xi_{k,n}^t$ are computed using the approximated posterior belief, which is given by

$$q^t(\mathbf{x}_k | \mathbf{y}) \propto I_{\Delta \to k}(\mathbf{x}_k) \prod_{n_r=1}^{N_r} \prod_{m \in V(k)} I_{m, n_r \to k}^{t-1}(x_{k,m}). \quad (21)$$

Then,

$$\mu_{k,n}^t = \sum_{\mathbf{a} \in \mathcal{X}_k} q^t(\mathbf{x}_k = \mathbf{a} | \mathbf{y}) a_n, \quad (22)$$

$$\xi_{k,n}^t = \sum_{\mathbf{a} \in \mathcal{X}_k} q^t(\mathbf{x}_k = \mathbf{a} | \mathbf{y}) |a_n - \mu_{k,n}^t|^2, \quad (23)$$

where $a_n$ is the $n$-th element of $N$-dimensional vector $\mathbf{a} \in \mathcal{X}_k$.

After termination, the posterior LLRs in (4) can now be calculated by

$$\Lambda(c_{k,j}) = \log \frac{\sum_{\mathbf{x}_k \in \mathcal{X}_{k,j}^+} q^t(\mathbf{x}_k | \mathbf{y})}{\sum_{\mathbf{x}_k \in \mathcal{X}_{k,j}^-} q^t(\mathbf{x}_k | \mathbf{y})}. \quad (24)$$

Summarizing the above discussion, the EPA decoding algorithm for SCMA is shown in Algorithm 1, where $N_{in}$ is the number of maximum inner iterations, e.g. 3, and $MAX$ is a large positive constant value, e.g. $MAX = 1000$.

### C. Complexity Analysis

In this subsection, the order of complexity for EPA receiver is analyzed. Table 1 gives the explanation of the key parameters, and Table 2 summarizes the complexity analysis results for different receiver types, with MMSE-SIC and MPA being the baselines. Note that the complexity order in Table 2 only includes the dominant terms, and the detailed number of calculations depends heavily on the choice of specific hardware implementations. It can be seen that the complexity order of EPA SCMA decoder scales only linearly with $M$ and $d_f$, as opposed to exponential order of complexity of MPA.

**Algorithm1: EPA SCMA Decoding**
- **Initialization**:
  $t = 1, \mu_{n,n_r \to k}^0 = 0, \xi_{n,n_r \to k}^0 = MAX, n = 1, \ldots, N, n_r = 1, \ldots, N_r, k \in F(n)$.
- **Iteration:** for t = $1: N_{in}$, do
  1) Compute $q^t(\mathbf{x}_k|\mathbf{y})$ via (21), $k = 1, \ldots, K$.
  2) Compute $\mu_{k,n}^t$ and $\xi_{k,n}^t$ via (22), (23), $k = 1, \ldots, K, n \in V(k)$.
  3) Compute $\xi_{k \to n, n_r}^t$ and $\mu_{k \to n, n_r}^t$ via (17), (18), $k = 1, \ldots, K, n \in V(k), n_r = 1, \ldots, N_r$.
  4) Compute $\mu_{n,n_r \to k}^t$ and $\xi_{n,n_r \to k}^t$ via (19), (20), $n = 1, \ldots, N, k \in F(n), n_r = 1, \ldots, N_r$.
- **LLR Calculation:**
  Compute $\Lambda(c_{k,j})$ and $\lambda_1(c_{k,j})$ via (24), (3), $k = 1, \ldots, K, j = 1, \ldots, J$.

Table 1: Key parameters in complexity analysis

| Parameters | Description |
|---|---|
| $N_r$ | Number of receiver antennas |
| $N$ | Spreading length of a spreading block |
| $K$ | Number of users/layers |
| $N_{iter}$ | Total number of iterations, including inner-loop and number of MPA/EPA rounds |
| $M$ | Codebook size |
| $M_p$ | Number of projection points on the constellation |
| $d_f$ | Degree of signal superposition on a given resource element |
| $d_s$ | Maximum degree-of-freedom allowed in SIC-MPA receiver |

Table 2: Complexity Comparison

| Receiver Type | Complexity Order (only dominant part considered) |
|---|---|
| MMSE-SIC (baseline) | $\mathcal{O}(N_r^3 N^3 K)$ |
| MPA (baseline) | $\mathcal{O}\left(N_{iter} N_r N M_p^{d_f}\right)$ |
| SIC-MPA | $\mathcal{O}\left(N_{iter} N_r N M_p^{d_s}\right)$ |
| EPA | $\mathcal{O}\left(N_{iter} N_r N M d_f\right)$ |

For ease of view, Table 3 presents the numerical examples to show the complexity comparison between different receivers for typical SCMA scenarios. Use of other schemes and the same receiver would yield similar trends and observations. For SIC-MPA receiver, both the cases of $d_s = 2$ and $d_s = 3$ are considered. Note that the total number of iterations $N_{iter} = N_{out} N_{in}$ for SIC-MPA is slightly larger than MPA and EPA. Typically, for inner loop iterations $N_{in} = 3$, the number of outer loop iterations $N_{out}$ is 3 for EPA/MPA ($N_{iter} = 9$) and 4 ($N_{iter} = 12$) for SIC-MPA. The percentage values in Table 3 represent the complexity ratios of different receivers to the baseline MMSE-SIC and MPA, respectively. For example, when $(M, M_p) = (8, 4)$, the values 14.1% (BL1) and 1.17% (BL2) corresponding to EPA indicate that the complexity ratio of EPA over baseline MMSE-SIC is 14.1%, and the complexity ratio of EPA over baseline MPA is 1.17%, respectively. As shown in Table 3, the complexity of the proposed EPA receiver is apparently lower than the full MPA and MMSE-SIC ($d_s = 3$). For $M = 4, 8$, the complexity of EPA is higher than SIC-MPA ($d_s = 2$), while for $M = 16$, EPA is slightly lower than SIC-MPA ($d_s = 2$).

Table 3: Complexity comparison for typical values of SCMA

| Parameter | $(N_r, N, K, N_{iter}, d_f) = (4,4,12,9(12),6)$ | | |
|---|---|---|---|
| | $(M, M_p)$ = (4,3) | $(M, M_p)$ = (8,4) | $(M, M_p)$ = (16,9) |
| MMSE-SIC (baseline1) | $\mathcal{O}(49152)$ | $\mathcal{O}(49152)$ | $\mathcal{O}(49152)$ |
| MPA (baseline2) | $\mathcal{O}(104976)$ | $\mathcal{O}(589824)$ | $\mathcal{O}(76527504)$ |
| SIC-MPA ($d_s = 3$) | $\mathcal{O}(5184)$ 10.5% (BL1) 4.94% (BL2) | $\mathcal{O}(12288)$ 25.0% (BL1) 2.08% (BL2) | $\mathcal{O}(139968)$ 285% (BL1) 0.18% (BL2) |
| SIC-MPA ($d_s = 2$) | $\mathcal{O}(1728)$ 3.52% (BL1) 1.65% (BL2) | $\mathcal{O}(3072)$ 6.25% (BL1) 0.52% (BL2) | $\mathcal{O}(15552)$ 31.6% (BL1) 0. 02% (BL2) |
| EPA | $\mathcal{O}(3456)$ 7.03% (BL1) 3.29% (BL2) | $\mathcal{O}(6912)$ 14.1% (BL1) 1.17% (BL2) | $\mathcal{O}(13824)$ 28.1% (BL1) 0.018% (BL2) |

IV. EXPERIMENTAL RESULTS

In this section, link level simulation of the proposed EPA based iterative SCMA receiver is performed. The codebook used in the simulation is $M = 8$ point codebook proposed in [13]. The channel model follows the TDL-A model [14] with delay spread 30ns and the moving speed is 3km/h. For channel coding, turbo code is adopted. The spectrum efficiency is defined as se = cr $*\log_2 M / N$ bps/Hz per user, where cr is the turbo code rate. The simulation results are averaged over 10,000 sub-frames. For comparison, the results of full MPA and SIC-MPA are given. In all the simulations, the numbers of inner loop iterations for the iterative receiver are set to be 3.

First, we examine the convergence of the EPA receiver. Fig. 3 shows BLER performance at different number of outer loop iterations for se = 0.2, $K = 12, N_r = 4$. It can be seen from Fig. 3 that the proposed EPA converges after 3 outer iterations. Similar results can be obtained for other scenarios, which demonstrate the good convergence of EPA receiver.

Fixing the number of outer loop iterations to be $N_{out} = 3$, the corresponding performance for se = 0.2, $K = 12, N_r = 4$ is shown in Fig. 4. The performances of both full MPA and SIC-MPA are given for comparison. Note that SIC-MPA has low convergence rate and $N_{out} = 4$ for SIC-MPA ($d_s = 2, 3$), and $N_{out} = 6$ for SIC-MPA ($d_s = 1$). As shown in Fig. 4, EPA achieves the same performance as MPA. For SIC-MPA, however, the value of $d_s$ affects the performance and in this case, $d_s = 3$ is good enough while there is some performance loss when $d_s = 1, 2$.

Finally, we evaluate the BLER performance of SCMA system with different spectrum efficiencies. Fig. 5 shows the BLER versus SNR under different spectrum efficiencies when there are $K = 12$ active users and $N_r = 4$ receive antennas. It can be seen from Fig. 5 that the BLER performance of the proposed EPA receiver is the same as MPA receiver for $se = 0.1 \sim 0.4$, while there is a slight performance loss for $se = 0.5$, e.g. about 0.2 dB loss at BLER $= 10^{-2}$.

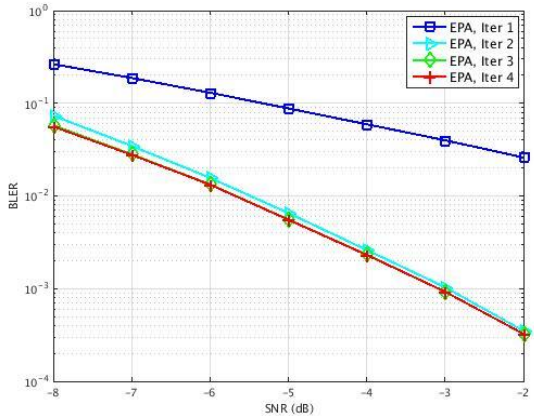

Fig. 3. Convergence performance of EPA when $K = 12, N_r = 4$, se = 0.2.

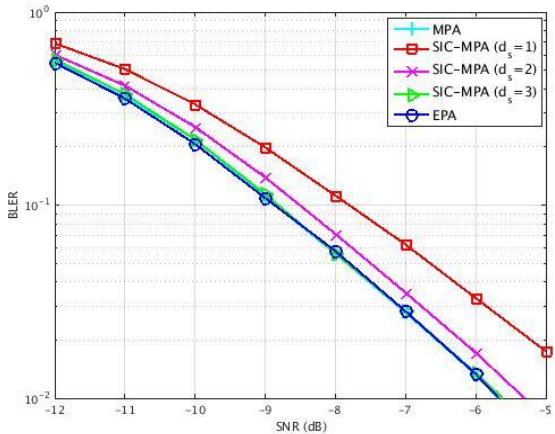

Fig. 4. BLER performance comparison for different receivers when $K = 12, N_r = 4$, se = 0.2.

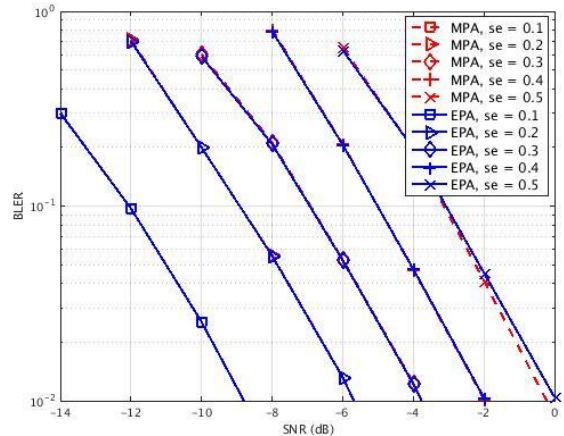

Fig. 5. BLER performance comparison at different spectrum efficiencies when $K = 12, N_r = 4$.

## V. CONCLUSION

In this paper, a low complexity iterative uplink SCMA receiver is proposed based on the expectation propagation algorithm (EPA). The complexity of proposed EPA receiver only scales linearly with the size of codebook and the degree of signal superposition on a given resource element. Link level simulation results demonstrate that the proposed EPA receiver has nearly the same BLER performance as full MPA receiver, which implies its potential use in practical system implementations.

ACKNOWLEDGMENT

This work is supported in part by National Science and Technology Major Project under grant No. 2015ZX03002010.

REFERENCES

[1] "5G: A Technology Vision," Huawei Technologies Co., Ltd., Shenzhen, China, Whitepaper Nov. 2013.
[2] H. Nikopour and H. Baligh, "Sparse code multiple access," in *Proc. IEEE 24th International Symposium on Personal Indoor and Mobile Radio Communications (PIMRC)*, pp. 332-336, 2013.
[3] M. Taherzadeh, H. Nikopour, A. Bayesteh and H. Baligh, "SCMA codebook design," in *Proc. IEEE Vehicular Technology Conference (VTC Fall)*, 2014, pp. 1-5.
[4] R. Hoshyar, F. P. Wathan and R. Tafazolli, "Novel low-density signature for synchronous CDMA systems over AWGN channel," *IEEE Trans. Signal Process.*, vol. 56, no. 4, pp. 1616-1626, Apr. 2008.
[5] G.D. Forney Jr, and L.F. Wei, "Multidimensional constellations. I.Introduction, figures of merit, and generalized cross nstellations,"*IEEE J. Sel. Areas Commun.*, vol. 7, no. 6, pp.877-892, 1989.
[6] F. R. Kschischang, B. J. Frey, and H. A. Loeliger, "Factor graphs and the sum-product algorithm," *IEEE Trans. Inf. Theory*, vol. 47, no. 2, pp. 498–519, Feb. 2001.
[7] A. Bayesteh, E. Yi, H. Nikopour. H. Baligh, "Low Complexity Techniques for SCMA Detection", in *IEEE Globecom Workshops*, San Diego, CA, 2015, pp.1-6.
[8] S. Zhang *et al.*, "Sparse code multiple access: An energy efficient uplink approach for 5G wireless systems," in *Proc. IEEE Global Commun. Conf.(GLOBECOM'14)*, Dec. 2014, pp. 4782–4787.
[9] R1-166098, "Discussion on the feasibility of advanced MU-detector," Huawei, HiSilicon, RAN1#86, Gothenburg, Sweden, Aug 22-26, 2016.
[10] T. P. Minka, "Expectation propagation for approximate Bayesian inference" in Uncertainty in Artificial Intelligence, 2001: 362-369.
[11] X. Meng, S. Wu, L. Kuang, Z. Ni and J. Lu, "Expectation Propagation Based Iterative Multi-User Detection for MIMO-IDMA Systems," *IEEE 79th Vehicular Technology Conference (Spring)*, Seoul, 2014, pp. 1-5.
[12] Y. Wu, S. Zhang and Y. Chen, "Iterative multiuser receiver in sparse code multiple access systems," *2015 IEEE International Conference on Communications (ICC)*, London, 2015, pp. 2918-2923.
[13] R1-164703, "LLS results for uplink multiple access," Huawei, HiSilicon, RAN1#85, Nanjing, China, May 16-20, 2016.
[14] 3GPP TR 38.900, v14.1.0, "Study on channel model for frequency spectrum above 6 GHz."